\documentclass[physrev,reprint,nofootinbib,superscriptaddress]{revtex4-2}

\pdfoutput=1

\usepackage{graphicx}
\usepackage{dcolumn}
\usepackage{bm}
\usepackage{amsmath}
\usepackage{amssymb}
\usepackage{booktabs}
\usepackage{multirow}
\usepackage{physics}
\usepackage{xcolor}
\usepackage{siunitx}
\usepackage{afterpage}
\usepackage{mleftright}

\usepackage{xtab}
\usepackage[multiple]{footmisc}
\usepackage{tcolorbox}

\usepackage{pdfpages}
\usepackage{etoolbox}

\usepackage[left=20mm,right=20mm,top=25mm,bottom=25mm,a4paper]{geometry}

\makeatletter
\patchcmd{\@outputpage@head}{\@ifx{\LS@rot\@undefined}{}{\LS@rot}}{}{}{}
\newcommand*{\balancecolsandclearpage}{%
  \close@column@grid
  \cleardoublepage
  \twocolumngrid
\newcommand\blankpage{%
  \null
  \thispagestyle{empty}%
  \addtocounter{page}{-1}%
  \newpage}
}
\makeatother

\protected\def\verythinspace{
  \ifmmode
    \mskip0.5\thinmuskip
  \else
    \ifhmode
      \kern0.08334em
    \fi
  \fi
}

\makeatletter
\renewcommand\@make@capt@title[2]{%
 \@ifx@empty\float@link{\@firstofone}{\expandafter\href\expandafter{\float@link}}%
  {\textbf{#1: }}#2\quad
}%
\makeatother

\usepackage{url}

\usepackage[colorlinks]{hyperref}
\hypersetup{%
        plainpages=true,
        breaklinks=true,
        hypertexnames=false,
        pageanchor=true,
        colorlinks=true,
        linkcolor={blue},
        citecolor={blue},
        urlcolor={blue},
        anchorcolor={black}
      }

\begin{document}

\title{Realisation of a Protected Cat-Qutrit Manifold via Engineered Quantum Tunnelling}

\author{Sangil Kwon}
\thanks{email: sangil.kwon@dgist.ac.kr}
\affiliation{Department of Physics and Chemistry, DGIST, Dalseong-gun, Daegu, 42988, South Korea}
\affiliation{Research Institute for Science and Technology, Tokyo University of Science, Shinjuku-ku, Tokyo, 162-8601, Japan}

\author{Daisuke Hoshi}
\affiliation{Department of Physics, Graduate School of Science, Tokyo University of Science, Shinjuku-ku, Tokyo, 162-8601, Japan}
\affiliation{RIKEN Center for Quantum Computing (RQC), Wako-shi, Saitama, 351-0198, Japan}

\author{Toshiaki Nagase}
\affiliation{Department of Physics, Graduate School of Science, Tokyo University of Science, Shinjuku-ku, Tokyo, 162-8601, Japan}
\affiliation{RIKEN Center for Quantum Computing (RQC), Wako-shi, Saitama, 351-0198, Japan}

\author{Daichi Sugiyama}
\affiliation{Department of Physics, Graduate School of Science, Tokyo University of Science, Shinjuku-ku, Tokyo, 162-8601, Japan}
\affiliation{RIKEN Center for Quantum Computing (RQC), Wako-shi, Saitama, 351-0198, Japan}

\author{Hiroto Mukai}
\affiliation{Department of Applied Physics, School of Engineering, The University of Tokyo, Bunkyo-ku, Tokyo, 113-8656, Japan}
\affiliation{RIKEN Center for Quantum Computing (RQC), Wako-shi, Saitama, 351-0198, Japan}
\affiliation{Research Institute for Science and Technology, Tokyo University of Science, Shinjuku-ku, Tokyo, 162-8601, Japan}

\author{Kengo Takemura}
\affiliation{Department of Physics, Graduate School of Science, Tokyo University of Science, Shinjuku-ku, Tokyo, 162-8601, Japan}
\affiliation{RIKEN Center for Quantum Computing (RQC), Wako-shi, Saitama, 351-0198, Japan}

\author{Rintaro Kojima}
\affiliation{Department of Physics, Graduate School of Science, Tokyo University of Science, Shinjuku-ku, Tokyo, 162-8601, Japan}
\affiliation{RIKEN Center for Quantum Computing (RQC), Wako-shi, Saitama, 351-0198, Japan}

\author{Yu Zhou}
\affiliation{RIKEN Center for Quantum Computing (RQC), Wako-shi, Saitama, 351-0198, Japan}

\author{Shohei Watabe}
\affiliation{College of Engineering, Shibaura Institute of Technology, Koto-ku, Tokyo, 135-8548, Japan}
\affiliation{Research Institute for Science and Technology, Tokyo University of Science, Shinjuku-ku, Tokyo, 162-8601, Japan}

\author{Fumiki Yoshihara}
\affiliation{Department of Physics, Graduate School of Science, Tokyo University of Science, Shinjuku-ku, Tokyo, 162-8601, Japan}
\affiliation{Research Institute for Science and Technology, Tokyo University of Science, Shinjuku-ku, Tokyo, 162-8601, Japan}

\author{Jaw-Shen Tsai}
\affiliation{Research Institute for Science and Technology, Tokyo University of Science, Shinjuku-ku, Tokyo, 162-8601, Japan}
\affiliation{Graduate School of Science, Tokyo University of Science, Shinjuku-ku, Tokyo, 162-8601, Japan}
\affiliation{RIKEN Center for Quantum Computing (RQC), Wako-shi, Saitama, 351-0198, Japan}


\date{\today}

\begin{abstract}
Engineering quantum tunnelling in phase space has emerged as a viable method for creating a protected logical qubit manifold with biased-noise properties.
A promising approach is to combine a Kerr nonlinearity with a multi-photon drive, resulting in a system known as a Kerr parametric oscillator (KPO).
In this work, we implement a three-photon KPO and explore its potential as a protected bosonic qutrit.
We confirm quantum coherence by demonstrating three-photon Rabi oscillations and performing direct Wigner function measurements that reveal the formation of three-component cat-like states.
Crucially, we observe a breathing-like dynamic in phase space, a characteristic feature of driven quantum systems.
This dynamic arises from macroscopic temporal interference between the cat-qutrit manifold and the excited states.
The frequency of resulting oscillations in the mean photon number provides a direct, time-domain measurement of the energy gap separating the qutrit from the excited states, thereby establishing an experimental hallmark of qutrit manifold protection.
Furthermore, we identify a parasitic higher-order pump term as the primary mechanism constraining the mean photon number, highlighting its mitigation as a requisite for maximising protection.
Our findings elucidate the basic quantum properties of the three-photon KPO and establish the first step towards its use as an alternative qutrit platform.
\end{abstract}

\maketitle


\section{Introduction}

Fault-tolerant quantum computing requires a platform where quantum information can be easily encoded and controlled while being robustly protected from various noise sources.
While superconducting circuits have emerged as a leading platform for quantum computation \cite{cQED, kwon, mit, gu}, their inherent susceptibility to noise, such as single-photon loss or dephasing, limits their path to full fault tolerance.
Quantum error correction (QEC) is expected to form the foundation of fault-tolerant quantum computing by resolving these issues;
however, the immense number of physical qubits required to construct even a single error-free logical qubit remains a significant barrier.

Bosonic codes address this challenge by leveraging the large Hilbert space of a resonator to encode a logical qubit into specialised quantum states \cite{terhal2020, cai2021, joshi2021, ma2021}.
These states are engineered to be correctable up to a certain error threshold, or at least robust against specific noise channels.
Nevertheless, because such states represent only a subset of the many states realisable in the resonator, they are highly susceptible to noise that causes leakage out of the computational subspace.
Therefore, robust preservation of logical qubit states against various noise sources is imperative,
as evidenced by the active \cite{campagne-ibarcq2020, sivak2023, brock2025} and autonomous \cite{lachance-quirion2024, sellem2025} stabilisation schemes required to prevent leakage in Gottesman--Kitaev--Preskill codes \cite{gkp}.

Within the framework of bosonic codes, a notable strategy for generating robust logical qubits is to implement a system where the logical qubit space is composed of energy eigenstates.
Kerr parametric oscillators (KPOs)---which induce quantum tunnelling through a combination of Kerr nonlinearity and a multi-photon drive \cite{dykman, goto2019b, wustmann2019}---are particularly well-suited for this approach.
In a two-photon KPO, even and odd cat states serve as the eigenstates spanning the logical qubit space \cite{wielinga1993, cochrane1999, marthaler2007, guo2013, minganti2016, goto2016a, goto2016b, puri2017a, zhang2017, catGen, bellCat}.
This manifold is protected against single-photon loss and low-frequency dephasing by an energy gap that separates the qubit space from the non-computational space.
This energy gap distinguishes KPO cat states from those generated via alternative methods, such as dynamic generation \cite{yurke1988, miranowicz1990, he2023} or interaction with a two-level system \cite{deleglise2008, vlastakis2013}.
In this context, KPO cat states inherently form a protected qubit manifold.

Furthermore, by adopting two opposite coherent states as the computational basis, the two-photon KPO exhibits a biased-noise property.
In this regime, bit-flip errors originating from single-photon loss are significantly suppressed compared to phase-flip errors \cite{puri2019, grimm2020, hajr2024, qing2025, frattini2024, venkatraman2024, ding2025}, analogous to cat states stabilised by dissipation engineering \cite{wolinsky1988, leghtas2015, lescanne2020, reglade2024}.
Together, the energy gap and the biased noise are expected to substantially reduce the resource overhead for quantum error correction \cite{darmawan2021}.

Another approach for reducing the hardware overhead is to use qudits \cite{lanyon2009, campbell2014, quditReview, KNF}.
Within superconducting platforms, most qudit implementations have relied on the energy levels of anharmonic oscillators, such as transmons \cite{blok2021, morvan2021, kononenko2021, goss2022, luo2023, roy2023, goss2024, wang2025, champion2025, zhou2025}.
However, these multi-level systems suffer from inherent transition asymmetries;
for instance, the $\ket{0} \rightarrow \ket{2}$ transition in a transmon is dipole-forbidden, complicating universal control.
Meanwhile, the implementation of qudits within the bosonic coding framework remains relatively unexplored (Ref.~\cite{brock2025} being a notable exception).
While multi-component cat states offer a high-dimensional alternative \cite{yurke1988, miranowicz1990, he2023, vlastakis2013}, they generally lack an inherent mechanism to protect against noise-induced leakage out of the computational space.
Furthermore, performing gate operations on such states is non-intuitive and requires more control overhead compared to multi-level systems.

Building on this research direction, the implementation of multi-component cat states within a protected manifold offers potential for reducing the resources required for quantum error correction.
A superconducting KPO provides an ideal platform for this approach;
for instance, three-component cat states can be prepared by engineering quantum tunnelling using modest Kerr nonlinearity and three-photon down-conversion \cite{braunstein1987, banaszek1997, felbinger1998, DSI, chang2020, eriksson2024}.
The symmetry of the states is enforced by the Hamiltonian itself, providing a more robust and controllable platform for high-dimensional quantum logic.
Previous demonstrations of the three-photon KPO operated in the semiclassical regime, and thus observed neither quantum coherence nor dynamics \cite{svensson2017, svensson2018}, despite numerous recent theoretical studies and proposals exploring its quantum nature \cite{zhang2017a, zhang2019, tadokoro2020, gosner2020, lang2021, arndt2022, miganti2023, iachello2023, guo2024, mora2024}.

In this work, we experimentally realise a three-photon KPO and demonstrate that the resulting quantum states constitute a protected qutrit manifold.
We emphasise that, in this context, protection denotes the suppression of population leakage into the non-computational subspace, enforced by the engineered energy gap.
The suppression of intra-manifold transitions between the three coherent-like states, i.e., the biased-noise property, remains a future milestone.

We first observe clear signatures of quantum coherence in the three-photon KPO, including three-photon Rabi oscillations and the characteristic interference pattern with three-fold symmetry (i.e., three-component cat-like states) in the measured Wigner functions.
Next, we investigate the quantum dynamics and relaxation of the prepared states, showing that the qutrit manifold is protected by an energy gap.
The size of this gap is measured via temporal interference between states inside and outside the qutrit manifold.
Finally, we identify the influence of higher-order pump terms as the primary physical mechanism suppressing the size of the KPO states compared to theoretical predictions for a purely three-photon driven system.

\section{Results}

\subsection{Setup}

\begin{figure*}
\centering
\includegraphics[scale=0.95]{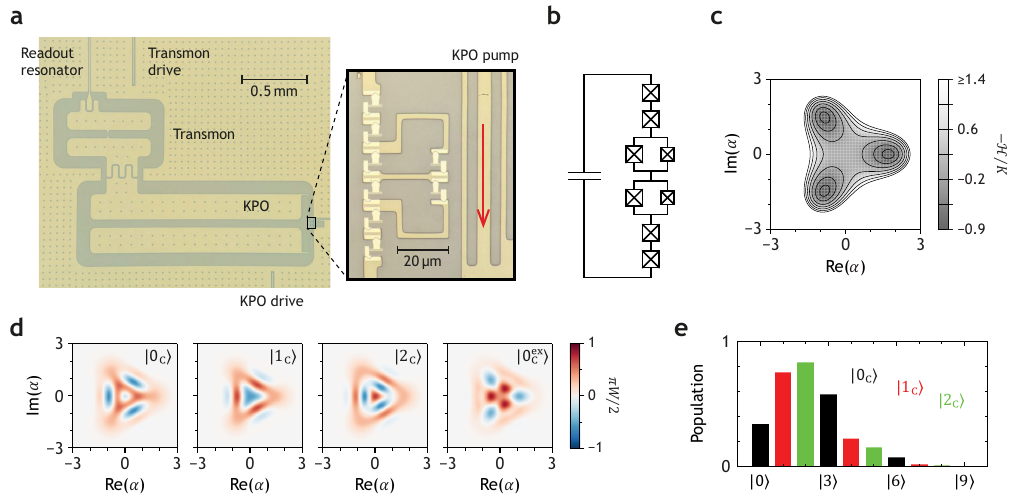}
\caption{\textbf{Chip and qutrit states.}
\textbf{a} Photograph of the chip containing the KPO.
The state of the KPO is measured by a nearby transmon and its readout resonator.
\textbf{b} Circuit diagram of the Kerr parametric oscillator (KPO), which consists of two direct-current superconducting quantum interference devices (DC SQUIDs) and four junctions with a shunting capacitor.
\textbf{c} The classical energy landscape corresponding to the Hamiltonian in Eq.~\eqref{eq:3PhKPO_cl}.
Following common practice, the inverted quasi-energy ($-\mathcal{H}/K$) is plotted such that the stable states appear as potential wells.
\textbf{d,e} Theoretical calculations of the Wigner functions and Fock-basis occupation probabilities for the eigenstates of the Hamiltonian in Eq.~\eqref{eq:3PhKPO}.
The states $\ket{0_\textrm{C}}$, $\ket{1_\textrm{C}}$, and $\ket{2_\textrm{C}}$ form the qutrit manifold;
$\ket{0_\textrm{C}^\textrm{ex}}$ is an excited state that is energetically closest to the qutrit manifold.
For (c)--(e), the parameters are $P/K = 0.822$, $\Delta/K = 0.72$, and $\eta = -0.05$.
}
\label{fig:chip}
\end{figure*}

The chip employed in this work is depicted in Fig.~\ref{fig:chip}a.
Our KPO design for this work (Fig.~\ref{fig:chip}b) was inspired by the recent work of NEC \cite{yamaguchi2024}.
A key distinction between the KPO used here and our previous designs \cite{catGen, bellCat} is the reduction in the number of direct-current superconducting quantum interference devices (DC SQUIDs), thereby minimising the flux modulation inhomogeneity.

The Hamiltonian of our system can be described as
(see Supplementary Information for the derivation)
\begin{equation}\label{eq:3PhKPO}
\begin{split}
\hat{\mathcal{H}} &= \Delta\, \hat{a}^\dagger\hat{a}
- \frac{K}{2} \hat{a}^\dagger\hat{a}^\dagger \hat{a}\hat{a} \\
&\quad\verythinspace
+ \frac{P}{2}
\mleft[ \mleft( \hat{a}^\dagger\hat{a}^\dagger\hat{a}^\dagger + \hat{a}\hat{a}\hat{a} \mright)
+ \eta \mleft( \hat{a}^\dagger\hat{a}^\dagger\hat{a}^\dagger\hat{a}^\dagger\hat{a}
+ \hat{a}^\dagger\hat{a}\hat{a}\hat{a}\hat{a} \mright) \mright].
\end{split}
\end{equation}
Here, we are working in units where $\hbar=1$;
$\hat{a}$ and $\hat{a}^\dagger$ are the ladder operators for the KPO;
$\Delta \equiv \omega_\textrm{K} - \omega_\textrm{p}/3$ is the KPO-pump frequency detuning, where
$\omega_\textrm{K}$ is the transition frequency between the $\ket{0}$ and $\ket{1}$ states, and $\omega_\textrm{p}$ is the frequency of the three-photon pump;
$K$ is the self-Kerr coefficient;
$P$ is the amplitude of the pump; and 
$\eta$ is the fractional strength of the higher-order pump term.
The case where $\eta = 0$ represents the idealised purely three-photon driven system, whereas a non-zero $\eta$ accounts for parasitic higher-order nonlinear processes that deviate from the ideal behaviour.
The Hamiltonian in Eq.~\eqref{eq:3PhKPO} is in the rotating frame defined by $\hat{\mathcal{H}}_0 = (\omega_\textrm{p}/3)\hat{a}^\dagger\hat{a}$.

The classical version of the Hamiltonian exhibits three stable states in phase space.
This classical Hamiltonian can be obtained by substituting $\hat{a}$ and $\hat{a}^\dagger$ in Eq.~\eqref{eq:3PhKPO} with the complex variables $\alpha$ and $\alpha^*$, respectively:
\begin{equation}\label{eq:3PhKPO_cl}
\begin{split}
\frac{\mathcal{H}}{K} &= \frac{\Delta}{K}(x^2+y^2) - \frac{1}{2}(x^2+y^2)^2 \\
&\quad\verythinspace
+\frac{P}{K}\mleft[\mleft(x^3-3xy^2 \mright) + \eta\mleft(x^5-2x^3y^2-3xy^4 \mright) \mright],
\end{split}
\end{equation}
where $x \equiv \textrm{Re}(\alpha)$ and $y \equiv \textrm{Im}(\alpha)$.

The consequence of this three-fold symmetry (Fig.~\ref{fig:chip}c) is that the eigenstates of a three-photon KPO are three-component cat-like states, as shown in Fig.~\ref{fig:chip}d.
In this work, the three nearly degenerate eigenstates closest to the energy minima form the qutrit manifold, $\ket{0_\textrm{C}}$, $\ket{1_\textrm{C}}$, and $\ket{2_\textrm{C}}$, which are distinguished by the remainder of their photon numbers modulo 3 (Fig.~\ref{fig:chip}e).
For example, in the $\ket{0_\textrm{C}}$ state, only Fock states with a photon number of $3n$ (where $n$ is a non-negative integer) are occupied.
Similarly, the excited state $\ket{0_\textrm{C}^\textrm{ex}}$ (the final plot in Fig.~\ref{fig:chip}d) also occupies only Fock states with $3n$ photons, similar to $\ket{0_\textrm{C}}$.

Throughout this work, the $P/K$ ratio is set to 0.822 (except in Fig.~\ref{fig:rabi}).
At this operating condition, the Kerr coefficient is determined to be $K/2\pi = 1.46$ MHz.
For reference, at zero pump amplitude, $\omega_\textrm{K}/2\pi = 3.112$ GHz and $K/2\pi = 1.72$ MHz.
These parameters were extracted from Rabi oscillation measurements that are explained in Sec.~\ref{sec:rabi}.
A more complete list of system parameters is presented in Supplementary Table~1.

\subsection{Three-photon Rabi oscillations}
\label{sec:rabi}

\begin{figure*}
\centering
\includegraphics[scale=0.95]{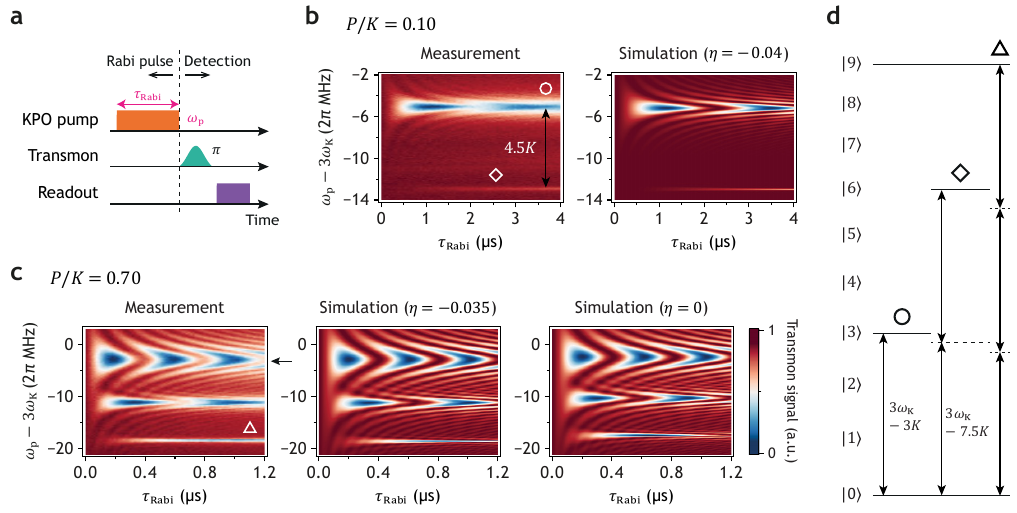}
\caption{\textbf{Three-photon Rabi oscillations.}
\textbf{a} Pulse sequence used for measuring Rabi oscillations. The control parameters are shown in magenta, and ``$\pi$'' denotes the transmon $\pi$-pulses.
\textbf{b,c} Rabi oscillations observed in the weak ($P \ll K$) and moderate ($P \sim K$) pump regimes, respectively.
In (b), the frequency axis begins from $-2$ MHz because the three-photon Rabi signal appears about $3K$ below $3\omega_\text{K}$.
The simulation results in (c) clearly show that the contribution of the higher-order pump term ($\eta$ term) must be included to quantitatively understand the Rabi oscillations.
Without $\eta$, the period of the higher-order signal (e.g., near $-10$ MHz) does not match well with the experimental data (see Methods section for more details).
Note that the experimental frequency axis ($\omega_\textrm{p} - 3\omega_\textrm{K}$) relates to the detuning via $\omega_\textrm{p} - 3\omega_\textrm{K} = -3\Delta$.
An arrow on the right side of the measurement panel in (c) indicates a specific detuning of $-3\Delta = 3.15$ MHz, which appears as $\Delta/K = 0.72$ in Fig.~\ref{fig:map}b.
The colour in both panels represents the population of the KPO's $\ket{0}$ state as detected by the transmon.
\textbf{d} Energy level diagram of the KPO. The symbols indicate the correspondence between the transitions and the observed signals.
}
\label{fig:rabi}
\end{figure*}

The validity of Eq.~\eqref{eq:3PhKPO} is confirmed by measuring Rabi oscillations in the population of the KPO $\ket{0}$ state.
It should be noted that these measurements demonstrate driven resonant transitions between discrete Fock states starting from the vacuum, serving to directly validate the multi-photon pump Hamiltonian before proceeding to the adiabatic preparation of the cat states.
Following the methodology detailed in Ref.~\cite{catGen}, the transmon exhibits photon number splitting once the KPO is excited, owing to the dispersive interaction between the two components.
By setting the frequency of the transmon $\pi$-pulse to match the transmon peak corresponding to the KPO $\ket{0}$ state, we can measure the population of this specific state.
The overall pulse sequence is shown in Fig.~\ref{fig:rabi}a.

Clear Rabi oscillations are observed at the pump frequency near $3\omega_\textrm{K}$, as shown in Fig.~\ref{fig:rabi}b and c.
The dynamics of the KPO are primarily governed by the ratio $P/K$.
In the weak pump regime ($P/K \ll 1$), two distinct signals are observed (Fig.~\ref{fig:rabi}b).
The Rabi oscillation denoted by the circle is induced by the transition between the $\ket{0}$ and $\ket{3}$ states (Fig.~\ref{fig:rabi}d) via a four-wave mixing process.
The weak signal, denoted by the diamond, originates from the transition between the $\ket{0}$ and $\ket{6}$ states via an eight-wave mixing process.
The frequency separation between these two signals is $4.5K$ because, in a Kerr nonlinear resonator, the energy spacing between $\ket{n}$ and $\ket{n+1}$ states is given by $\omega_\textrm{K} - nK$.
As we increase $P$, the Rabi oscillation pattern becomes more complicated, and additional higher-order transitions are involved.
The signal denoted by the triangle in Fig.~\ref{fig:rabi}c, for example, roughly corresponds to the transition between the $\ket{0}$ and $\ket{9}$ states (Fig.~\ref{fig:rabi}d).
All data presented in Fig.~\ref{fig:rabi} show excellent agreement with the simulation results, thus validating the use of Eq.~\eqref{eq:3PhKPO} as the model for our KPO.
For further details regarding the Hamiltonian characterisation procedure, please see Methods section.
(For all simulations in this work, QuTiP was used \cite{qutip}.)

The fitted value of $\eta$ is $-0.045 \pm 0.006$
(where the error denotes the standard deviation across datasets with various pump amplitudes),
which is significantly larger in magnitude than the theoretical value of approximately $-0.001$
(from Supplementary Eq.~(3)).
The reason for such a large fitted $\eta$ is unclear.
Up to $P/K=1.0$, $\eta$ does not show a systematic dependence on the pump amplitude (see Supplementary Fig.~2b).
This lack of dependence suggests that the unexpectedly large $\eta$ is likely related to unaccounted circuit degrees of freedom---such as individual junction degrees of freedom, given that our theoretical calculation of $\eta$ assumed a single degree of freedom (see Supplementary Information)---rather than heating effects triggered by a strong pump.

To address the uniqueness of $\eta$, we also considered other higher-order mechanisms,
including a sixth-order Kerr term
($\hat{a}^\dagger \hat{a}^\dagger \hat{a}^\dagger \hat{a}\hat{a}\hat{a}$)
and a six-photon pump
($\hat{a}^\dagger \hat{a}^\dagger \hat{a}^\dagger \hat{a}^\dagger \hat{a}^\dagger \hat{a}^\dagger
+ \hat{a}\hat{a}\hat{a}\hat{a}\hat{a}\hat{a}$).
However, these alternative models failed to reproduce the frequencies and temporal periods of the observed higher-order Rabi signals (e.g., the signals labelled with diamond and triangle markers in Fig.~\ref{fig:rabi}).
Therefore, we conclude that the $\eta$ term provides the simplest and most reliable empirical description of the system dynamics among the considered models.

\subsection{Quantum coherence in Wigner functions}
\label{sec:quantum}

\begin{figure*}
\centering
\includegraphics[scale=0.95]{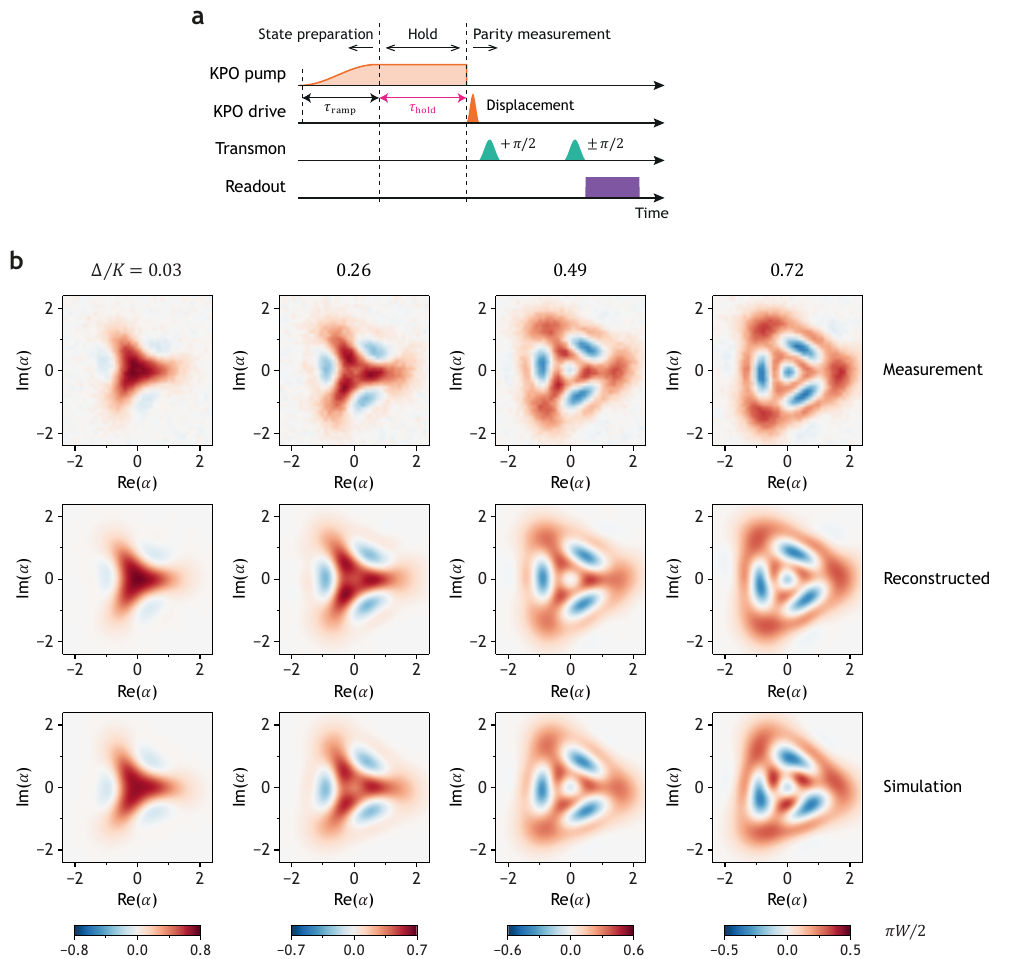}
\caption{\textbf{Wigner functions of three-photon KPO states.}
\textbf{a} Pulse sequence for measuring the Wigner function.
The pump-pulse conditions are as follows:
$P/K = 0.822$, $P_\textrm{CD}/P = -0.3$, where $P_\textrm{CD}$ is the amplitude of the counterdiabatic component of the pump, $\tau_\textrm{ramp} = 0.4$ \unit{\micro\second}, and $\tau_\textrm{hold} = 0.1$ \unit{\micro\second}.
\textbf{b} Wigner functions of the KPO states for various $\Delta/K$ values.
The measured raw data are shown in the top row, the corresponding reconstructed Wigner functions in the middle row, and the Wigner functions of the simulated states in the bottom row.
For the simulations, $\eta = -0.05$ was used and $T_1$ was assumed to be 3.5 \unit{\micro\second} (see Fig.~\ref{fig:relax}).
From left to right (corresponding to the individual panels), the mean photon numbers extracted from density matrices reconstructed from the experimental data are [0.58, 1.22, 2.20, 2.61].
The corresponding mean photon numbers obtained from the simulated density matrices are [0.61, 1.36, 2.33, 2.49].
The fidelities between the reconstructed and simulated states are [0.98, 0.95, 0.95, 0.95].
The fidelities between the reconstructed and $\ket{0_\text{C}}$ states (with $\eta = -0.05$) are [0.90, 0.87, 0.66, 0.56].
}
\label{fig:map}
\end{figure*}

We access the $\ket{0_\textrm{C}}$ state by ramping up the pump pulse with a $\sin^2(\pi t/2\tau_\textrm{ramp})$ profile, following the methodology presented in Refs.~\cite{catGen,bellCat}.
The resulting state of the KPO is then characterised by measuring its Wigner function via parity measurement \cite{royer1977, lutterbach1997}.
The pulse sequence for this process is shown in Fig.~\ref{fig:map}a.
While this ramping profile is conventionally used for adiabatic preparation, the inclusion of a negative counter-diabatic pulse ($P_\textrm{CD}/P = -0.3$) purposefully introduces a finite degree of non-adiabaticity.
This induces a small occupation of the excited states, the consequences of which are discussed in Sec.~\ref{sec:relax}.

The Wigner functions of our three-photon KPO states exhibit two essential features (Fig.~\ref{fig:map}b).
First, all data show triangular symmetry, which indicates the three-photon nature of our pump.
Secondly, the interference-like pattern---a clear signature of quantum coherence---appears and becomes more evident as the pump detuning ($\Delta$) increases, confirming that the quantum state of our KPO is a three-component cat-like state.
The dependence of the cat-state size on $\Delta$ is examined in greater detail in Sec.~\ref{sec:steady}.

The fidelity between the reconstructed and simulated density matrices is reasonably good, consistently $\gtrsim\! 0.95$ (see the caption of Fig.~\ref{fig:map}b).
Here, the reconstructed density matrices of the KPO were obtained via quantum state tomography with conditional generative adversarial network (QST-CGAN) \cite{ahmed2021a, ahmed2021b}, followed by the correction of unwanted Kerr evolution during the displacement pulse \cite{catGen}.
Further details regarding the density matrix reconstruction are provided in the Methods section.

To compare our measured Wigner function with simulation, we must account for the effect of single-photon loss during the parity measurement.
This is because the parity measurement time ($\approx\! 0.67$ \unit{\micro\second}) is not negligible compared to the $T_1$ of our KPO ($\approx\! 4.5$ \unit{\micro\second}).
We found that the Wigner functions calculated using an effective $T_1$ ($T_{\text{1eff}}$) that is twice the KPO's actual $T_1$ show good agreement with the parity measurement simulation with a three-component cat state (see Supplementary Fig.~3).
This suggests that the effects of loss in the KPO during the parity measurement can be captured by assuming $T_{\text{1eff}}=2T_1$.
Here, the factor of 2 is an empirical coincidence specific to the parameters used in this particular simulation.
All simulations presented in Fig.~\ref{fig:map}b include the effect of single-photon loss using $T_1$ for the state preparation and hold processes, and using $T_{\text{1eff}}$($=2T_1$) for the parity measurement.

\subsection{Dynamics and relaxation}
\label{sec:relax}

\begin{figure*}
\centering
\includegraphics[scale=0.95]{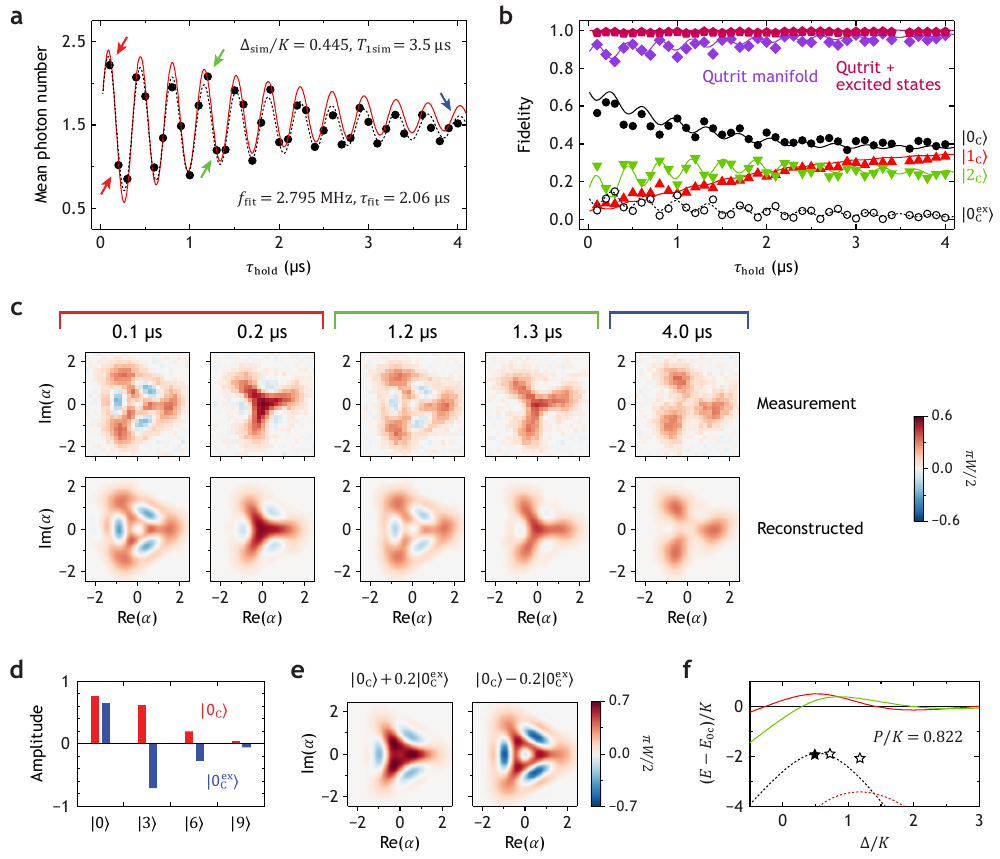}
\caption{\textbf{Dynamics and relaxation.}
\textbf{a} Time evolution of the mean photon number, showing oscillations caused by imperfect adiabatic conversion from the $\ket{0}$ state to the $\ket{0_\textrm{C}}$ state.
The dashed line is a fit to a simple decaying cosine function, with fitted parameters $f_\text{fit}$ and $\tau_\text{fit}$ shown.
\textbf{b} Time evolution of the fidelity for the qutrit states and $\ket{0_\textrm{C}^\textrm{ex}}$.
Within the label ``Qutrit + excited states,'' the excited states refer to the subspace spanned by the states $\ket{0_\textrm{C}^\textrm{ex}}$, $\ket{1_\textrm{C}^\textrm{ex}}$, and $\ket{2_\textrm{C}^\textrm{ex}}$.
In both (a) and (b), the simulation results (solid lines) account for single-photon loss in the KPO during parity measurement using an effective time $T_{\text{1eff}}=2T_\textrm{1sim}$.
The parameters used for the simulation, $\Delta_\text{sim}$ and $T_\text{1sim}$, are provided in (a).
\textbf{c} Wigner functions of the KPO state at the times indicated by arrows in (a).
\textbf{d} The amplitudes of the Fock basis for the $\ket{0_\textrm{C}}$ and $\ket{0_\textrm{C}^\textrm{ex}}$ states.
\textbf{e} Superpositions of the states $\ket{0_\textrm{C}}$ and $\ket{0_\textrm{C}^\textrm{ex}}$ with two opposing relative phases.
The mean photon number is
$0.78$ for the state $\ket{0_\textrm{C}}+0.2\ket{0_\textrm{C}^\textrm{ex}}$ and 
$2.05$ for $\ket{0_\textrm{C}}-0.2\ket{0_\textrm{C}^\textrm{ex}}$.
For (d) and (e), $\Delta/K = 0.445$.
\textbf{f} The first five quasienergy levels $E$ calculated using Eq.~\eqref{eq:3PhKPO}.
The solid lines indicate the quasienergy levels for the qutrit states:
black ($\ket{0_\textrm{C}}$, $E_\textrm{0c}$), red ($\ket{1_\textrm{C}}$), and green ($\ket{2_\textrm{C}}$).
The dashed lines show the quasienergy levels for the excited states: black ($\ket{0_\textrm{C}^\textrm{ex}}$) and red ($\ket{1_\textrm{C}^\textrm{ex}}$).
The stars represent the oscillation frequencies extracted from the mean photon number data;
the black solid star specifically corresponds to $f_\text{fit}$ in (a).
Here, the energy levels of the excited states are lower than those of the qutrit states, a consequence of the negative sign of $K$ in Eq.~\eqref{eq:3PhKPO}.
For all simulations in this figure, $P/K = 0.822$, $P_\textrm{CD}/P = -0.3$, and $\eta = -0.05$.
}
\label{fig:relax}
\end{figure*}

Now we explore the quantum dynamics of the $\ket{0_\textrm{C}}$ state, which are associated with the occupation of the excited state $\ket{0_\textrm{C}^\textrm{ex}}$ and single-photon loss.
We first prepared the $\ket{0_\textrm{C}}$ state following the procedure in Fig.~\ref{fig:map}a and subsequently measured its Wigner functions at various hold times ($\tau_\textrm{hold}$).
From this set of Wigner functions, we extracted the density matrices to calculate the mean photon numbers (Fig.~\ref{fig:relax}a) and the fidelities of both the qutrit and excited states (Fig.~\ref{fig:relax}b).

We observed a strong oscillation in the mean photon number.
This oscillation is a consequence of a breathing-like dynamic---a periodic expansion and contraction of the KPO state in phase space---as clearly shown by the Wigner functions in Fig.~\ref{fig:relax}c.
This dynamic is primarily caused by the oscillation of the relative phase between the $\ket{0_\textrm{C}}$ and $\ket{0_\textrm{C}^\textrm{ex}}$ states, rather than by population transfer.
The populations of these states, shown in Fig.~\ref{fig:relax}b, were directly extracted from the reconstructed density matrices.
This confirms that the observed dynamics result from the coherent involvement of the first excited state $\ket{0_\textrm{C}^\textrm{ex}}$, whose population is quantitatively monitored throughout the process.
This superposition is induced by the finite adiabaticity of the state preparation process, enabling the characterisation of the energy gap separating the qutrit and excited manifolds.

We claim that this exotic breathing dynamic is a manifestation of macroscopic temporal interference,
as it involves the interference of multiple photon-number states.
Such a phenomenon is characteristic of driven quantum systems.
In an un-driven system, a superposition of eigenstates maintains a constant mean photon number.
Even within this driven KPO, interference between qutrit states (e.g., $\ket{0_\textrm{C}}$ and $\ket{1_\textrm{C}}$) yields no such breathing.
To elucidate this, consider the expectation value of the photon number operator,
$\hat{N} = \hat{a}^\dagger \hat{a}$,
for a superposition $\ket{\psi(t)} = c_1(t)\!\ket{\psi_1} + c_2(t)\!\ket{\psi_2}$,
where $\ket{\psi_1}$ and $\ket{\psi_2}$ are arbitrary quantum states:
\begin{equation}\label{eq:oscillation}
\begin{split}
\langle \hat{N} \rangle 
&= \left|c_1\right|^2 \!\mel{\psi_1}{\hat{N}}{\psi_1} 
+ \left|c_2\right|^2 \!\mel{\psi_2}{\hat{N}}{\psi_2} \\
&\quad
+ \left( c_1^* c_2 \mel{\psi_1}{\hat{N}}{\psi_2} 
+ c_1 c_2^* \mel{\psi_2}{\hat{N}}{\psi_1} \right).
\end{split}
\end{equation}

In conventional qubits, the cross terms vanish due to the orthogonality between qubit states.
Similarly, in the qutrit space of a three-photon KPO, the cross terms, such as $\mel{0_\textrm{C}}{\hat{N}}{1_\textrm{C}}$, also vanish because $\hat{N}$ does not induce transitions between different Fock number bases;
the state $\hat{N}\ket{0_\textrm{C}}$ remains within the Fock subspace of $3n$ photons,
whereas $\hat{N}\ket{1_\textrm{C}}$ occupies only the $3n+1$ subspace.
This absence of shared Fock states forbids any interference-driven oscillation in the mean photon number.

In contrast, the excited state $\ket{0_\textrm{C}^\textrm{ex}}$ shares the same modulo-3 symmetry (occupying $3n$ Fock states) with $\ket{0_\textrm{C}}$.
The oscillations in the mean photon number primarily rely on the cross terms in Eq.~\eqref{eq:oscillation} rather than the state populations.
The mean photon numbers of the $\ket{0_\textrm{C}}$ and $\ket{0_\textrm{C}^\textrm{ex}}$ states are 1.39 and 1.99, respectively.
Since the population of the $\ket{0_\textrm{C}^\textrm{ex}}$ state remains $<0.2$ at all times in Fig.~\ref{fig:relax}b, the maximum possible change in the mean photon number caused by population oscillations is $<\!\left|1.99-1.39 \right| \times 0.2=0.12$.

Note that the amplitudes of the $\ket{3n}$ Fock states have opposite signs between $\ket{0_\textrm{C}}$ and $\ket{0_\textrm{C}^\textrm{ex}}$, with the exception of the $\ket{0}$ state, as illustrated in Fig.~\ref{fig:relax}d.
The resulting non-zero cross term ($\!\mel{0_\textrm{C}}{\hat{N}}{0_\textrm{C}^\textrm{ex}} \neq 0$) explains how a small participation from $\ket{0_\textrm{C}^\textrm{ex}}$ ($<\!4$\% in population) can lead to mean photon number fluctuations greater than $1$ (Fig.~\ref{fig:relax}e).
Due to the energy gap $\mathcal{E}_\text{g}$ between these states, this interference term evolves proportionally to $\cos(\mathcal{E}_\text{g}t/\hbar)$, providing a direct means to extract the energy gap from the oscillation frequency.
We measured the energy gap at three different values of $\Delta/K$, represented by the stars in Fig.~\ref{fig:relax}f.
At $\Delta/K=1.114$, a notable difference is observed between the measured energy gap and the theoretical calculation.
While this discrepancy might arise from higher-order terms that are not captured in Eq.~\eqref{eq:3PhKPO}, identifying its exact physical origin remains an open question for future investigation.

We note that such interference is a generic feature of any KPO.
An $n$-photon KPO ($n \geq 2$) would exhibit similar behaviour arising from the superposition between the logical manifold and the excited states.
For instance, we believe the mean photon number oscillations observed in Fig.~4a of Ref.~\cite{wang2019} arise from a similar interference phenomenon.
Consequently, such breathing dynamics serve as a highly sensitive, in-situ indicator for population leakage out of the target logical manifold.

Two key consequences arise from the energy gap between the qutrit manifold and the excited states.
First, the total population of the qutrit manifold (represented by the purple diamond symbols in Fig.~\ref{fig:relax}b) exhibits a recovery towards 1, instead of a decay, apart from minor oscillations induced by imperfect adiabatic state preparation.
This is a clear manifestation of the protection of the qutrit space from low-frequency dephasing noise.
Such a conclusion is further supported by the good agreement between the measurement and the simulation without dephasing (solid lines in the same figure).

A simulation incorporating white-noise dephasing, with a timescale $T_\varphi$ inferred from $T_1$ and $T_2$, is presented in Supplementary Fig.~4a.
Unlike the results in Fig.~\ref{fig:relax}b, the simulation with dephasing shows the $\ket{0_\textrm{C}^\textrm{ex}}$ state population reaching a steady-state value of 0.126, which contradicts the experimentally observed population of 0.01 for the $\ket{0_\textrm{C}^\textrm{ex}}$ state at $\tau_\text{hold}=4$ \unit{\micro\second} (black open circles in Fig.~\ref{fig:relax}b).
Moreover, in Supplementary Fig.~4a, the population of the qutrit manifold decays (purple solid line), contrary to the recovery in Fig.~\ref{fig:relax}b and Supplementary Fig.~4b.

More quantitatively, by fitting the baseline of the $\ket{0_\textrm{C}^\textrm{ex}}$ state population data in Fig.~\ref{fig:relax}b with an exponential decay function, we extract a decay time of $2.53 \pm 0.44$ \unit{\micro\second}, where the uncertainty represents one standard deviation.
This timescale can be considered the manifold confinement time.
A similar fitting procedure applied to the qutrit manifold population yields a recovery time of $2.78 \pm 0.68$ \unit{\micro\second}, which is consistent with the $\ket{0_\textrm{C}^\textrm{ex}}$ state result.
In contrast, the intra-manifold decay time, such as that of the $\ket{0_\textrm{C}}$ state extracted from the same fitting, is $1.36 \pm 0.20$ \unit{\micro\second}, which is clearly different from the manifold confinement time.
The sensitivity of these population measurements is approximately $0.01$, as determined by the root-mean-square error of the fit residuals.

We stress that this energy gap protects against population leakage caused by dephasing noise with little spectral weight at or above the gap frequency ($\approx\,$2.8 \unit{\mega\hertz} in this case)---it does not suppress population leakage induced by photon loss or high-frequency dephasing noise.

Second, single-photon loss drives a cyclic population transfer within the qutrit manifold
($\ket{0_\textrm{C}} \rightarrow \ket{2_\textrm{C}} \rightarrow \ket{1_\textrm{C}} \rightarrow \ket{0_\textrm{C}} \rightarrow \cdots$), which is distinct from the standard ladder-like decay observed in conventional transmon-based qutrits \cite{morvan2021}.
Such relaxation originates directly from the underlying Fock-state composition.
Because the $\ket{0_\textrm{C}}$, $\ket{1_\textrm{C}}$, and $\ket{2_\textrm{C}}$ states are exclusively composed of Fock states with photon numbers $3n$, $3n+1$, and $3n+2$, respectively,
the loss of a single photon (the application of the annihilation operator $\hat{a}$) deterministically shifts the photon number modulo 3.
Consequently, $3n \rightarrow 3n-1 (\equiv 2 \pmod 3)$, dictating the observed transition sequence.

The data presented in Fig.~\ref{fig:relax}b clearly illustrate this sequential process:
as $\tau_\textrm{hold}$ increases, the fidelity of the $\ket{0_\textrm{C}}$ state decays.
This depletion is accompanied by an initial rise in the $\ket{2_\textrm{C}}$ state fidelity,
which subsequently feeds into the $\ket{1_\textrm{C}}$ state population.
As the system approaches the steady state, the populations of the three qutrit states converge towards a nearly equal distribution.

Notably, the fidelities of the $\ket{0_\textrm{C}}$ and $\ket{2_\textrm{C}}$ states in Fig.~\ref{fig:relax}b oscillate out of phase.
Simulations reproduce the oscillations observed in the $\ket{2_\textrm{C}}$ state, indicating that this dynamic primarily stems from the enhanced relaxation rate of states with larger mean photon numbers during the parity measurement (compare Fig.~\ref{fig:relax}a and b).

Lastly, in Fig.~\ref{fig:relax}, we use $T_\text{1sim} = 3.5$ \unit{\micro\second} as a fitting parameter, which is slightly shorter than the independently measured average value of $4.5$ \unit{\micro\second}. 
This discrepancy can be attributed to the typical temporal fluctuations of the relaxation time, as well as a possible reduction in $T_1$ under the pump;
the pump-induced AC Stark-like shift may bring the transmon frequency closer to weakly coupled two-level system defects.

\subsection{Steady states}
\label{sec:steady}

\begin{figure}
\centering
\includegraphics[scale=0.95]{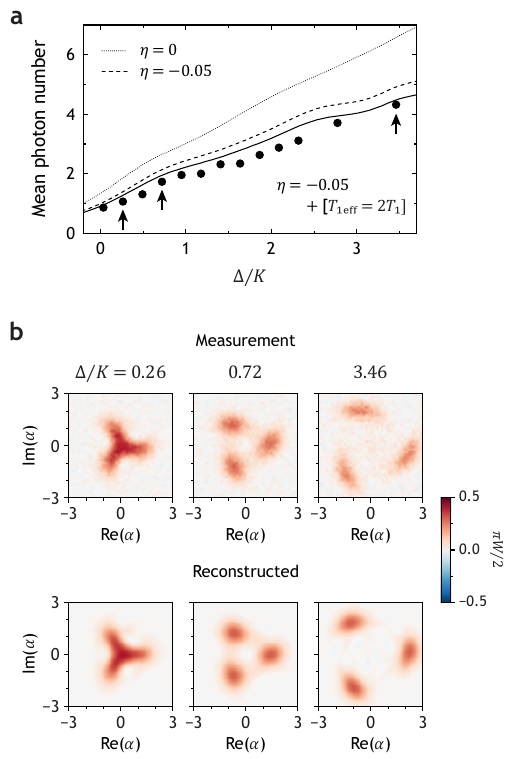}
\caption{\textbf{Steady states.}
\textbf{a} The mean photon number of the steady states as a function of the detuning, $\Delta/K$.
The symbols represent the values extracted from the density matrices reconstructed from the measured Wigner functions.
The lines are results from the steady-state solution of the Lindblad master equation derived from Eq.~\eqref{eq:3PhKPO}.
The effect of single-photon loss during the parity measurement is included in the solid line by using a relaxation time of $T_1 = 3.5$ \unit{\micro\second}.
The pump-pulse conditions are as follows:
$P/K = 0.822$, $P_\textrm{CD}/P = 0$, $\tau_\textrm{ramp} = 15$ \unit{\micro\second}, and $\tau_\textrm{hold} = 0.2$ \unit{\micro\second}.
A long $\tau_\textrm{ramp}$ was chosen instead of a long $\tau_\textrm{hold}$ to eliminate the occupation of excited states and, consequently, to avoid the complex dynamics shown in Fig.~\ref{fig:relax}.
\textbf{b} Wigner functions of the steady state at the $\Delta/K$ values indicated by the arrows in (a).
}
\label{fig:steady}
\end{figure}

It is well established that both the biased-noise profile and the protection of the qutrit manifold are enhanced as the mean photon number (i.e., the size of the cat states) increases \cite{puri2019, grimm2020, frattini2024, venkatraman2024, hajr2024, qing2025}.
Thus, to demonstrate the ability to achieve large cat states in this system, we measured the steady-state mean photon number as a function of the pump detuning $\Delta/K$ (Fig.~\ref{fig:steady}a).
As dictated by the classical Hamiltonian (Eq.~\eqref{eq:3PhKPO_cl}), the distance of the potential minima from the origin in phase space scales roughly as $\sqrt{\Delta/K}$.
Consequently, as the detuning increases, the potential wells are pushed further outwards, resulting in cat states with larger mean photon numbers (Fig.~\ref{fig:steady}b).

The measured mean photon number increases significantly more slowly than expected from a pure three-photon KPO model (Eq.~\eqref{eq:3PhKPO} with $\eta=0$, dotted line in Fig.~\ref{fig:steady}a).
We identify that the higher-order pump term with $\eta=-0.05$ (dashed line) is primarily responsible for this deviation from the idealised model.
Another mechanism that must be considered is single-photon loss during the parity measurement.
This is relevant because the pump is turned off during the parity measurement, causing the protective energy gap to vanish.
Consequently, in the absence of this protection, single-photon loss causes the state to decay towards the vacuum, reducing the size of the cat state.
By additionally incorporating this loss mechanism, the final simulation result (solid line) yields a reasonably good agreement with the experimental data.

\section{Discussion}

In this work, we investigated the quantum properties of a three-photon KPO, validating its potential as a protected qutrit platform.
We demonstrated quantum coherence by measuring three-photon Rabi oscillations and the characteristic interference patterns with three-fold symmetry (i.e., three-component cat-like states) in Wigner functions.
Analysis of the system's time evolution revealed two crucial dynamic features associated with qutrit manifold protection.
First, the oscillation in the mean photon number, caused by temporal interference between the qutrit and excited states, indicates the existence of the energy gap separating the qutrit manifold from the excited states.
The accompanying breathing dynamic in phase space is characteristic of driven quantum systems.
Second, we observed a distinct cyclic population transfer induced by single-photon loss.
Furthermore, our investigation of the steady-state mean photon number as a function of the pump detuning revealed the importance of mitigating the higher-order pump term for maximising the qutrit manifold protection.

Future work will focus on exploring the inherent biased-noise properties of this system.
A key component for established biased-noise systems is the ability to prepare states autonomously localised in phase space, similar to the coherent states generated in a two-photon KPO.
The Wigner functions shown in Fig.~\ref{fig:steady}b suggest that such localised states can be prepared around and above $\Delta/K = 0.5$ when $P/K = 0.822$.
Here, we define localisation as the regime where the three constituent coherent lobes are sufficiently separated in phase space such that their overlap is negligible, which is visually evidenced by the emergence of distinct peaks and clear central interference fringes.
Our previous work \cite{catGen} generated coherent states by preparing the $\ket{0}+\beta \ket{1}$ state in the Fock basis (where $\beta$ is a phase factor) and subsequently converting this state to the cat state basis.
However, this method is difficult in the three-photon KPO because it requires preparing a state of the form $\ket{0}+\beta \ket{1}+\gamma \ket{2}$, making the task of setting the proper phase factors, $\beta$ and $\gamma$, challenging.
A more convenient approach for preparing the desired localised state would be measurement-based state preparation, as previously demonstrated in the two-photon KPO \cite{frattini2024, venkatraman2024, hajr2024, qing2025}.

\begin{figure*}
\centering
\includegraphics[scale=0.95]{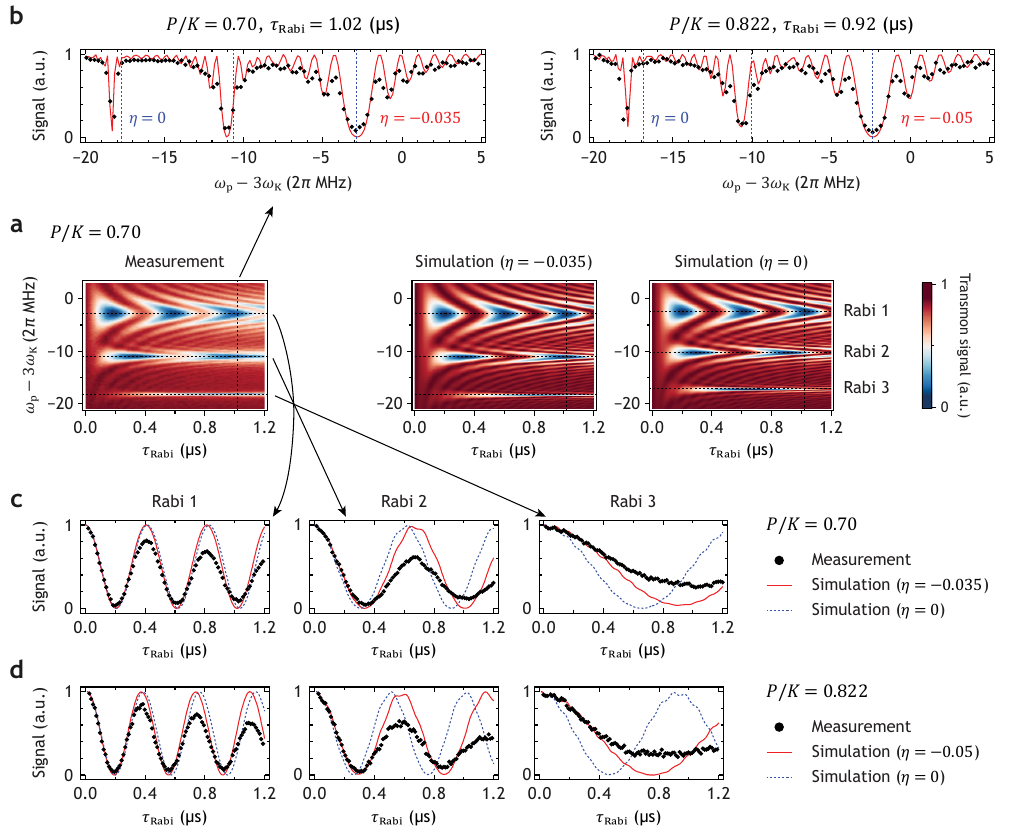}
\caption{\textbf{Fitting of three-photon Rabi oscillations.}
\textbf{a} Three-photon Rabi oscillations corresponding to those shown in Fig.~\ref{fig:rabi}c.
\textbf{b} Frequency slices from the full Rabi oscillation data.
The left panel corresponds to a slice from (a), whilst the right panel displays a similar measurement at a different $P/K$ ratio.
Black symbols represent the measured data;
solid red lines indicate simulations based on Eq.~(1) in the main text with a non-zero $\eta$;
and vertical blue dotted lines denote the frequency positions of the three Rabi signals extracted from similar simulations but with $\eta = 0$.
For ease of comparison, the position of the rightmost dotted line, which indicates the Rabi 1 position, is adjusted to match the Rabi 1 position extracted from the simulation with a non-zero $\eta$ (solid red line).
\textbf{c,d} Three representative Rabi oscillation traces at $P/K=0.70$ (c) and $P/K=0.822$ (d).
The simulations do not include relaxation or dephasing.
}
\label{fig:fitting}
\end{figure*}

One potential difficulty for observing the biased-noise properties is that any possible superposition between the qutrit states would be nonstationary.
This is because the qutrit states are never energetically degenerate simultaneously unless $P/K$ or $\Delta/K$ becomes very large (Fig.~\ref{fig:relax}f).
In comparison, the two-photon KPO possesses stationary qubit states when $\Delta/K$ is a non-negative integer \cite{catGen}.
Since the dynamics of nonstationary qutrit states are unitary, they are reversible in principle.
Thus, finding convenient means for generating an effective stationary state will be necessary.

The next crucial experimental milestones involve demonstrating operation of a universal qutrit gate set.
Furthermore, extending the platform to implement a protected ququart ($d=4$) represents another fascinating research direction, particularly given its potential for realising autonomous quantum error correction \cite{kwon2022}.
Finally, we suggest that a deeper theoretical and experimental investigation into how tunnelling dynamics \cite{reynoso2025, su2025} contribute to relaxation processes in such qudit systems would yield valuable insights for minimising decoherence across parametrically driven quantum processors.

\section{Methods}

\subsection{Characterisation of the Hamiltonian}
\label{sec:calibration}

We extract the pump amplitude ($P$), Kerr coefficient ($K$), higher-order pump parameter ($\eta$), and KPO transition frequency ($3\omega_\mathrm{K}$) by comparing the two-dimensional plot of Rabi oscillations with simulations based on Eq.~(1) in the main text.
(Characterisation of $3\omega_\mathrm{K}$ is required since it shifts as a function of the pump amplitude owing to an AC Stark-like shift.)
However, acquiring high-resolution two-dimensional Rabi oscillation data, such as that shown in Fig.~\ref{fig:fitting}a, is a time-consuming process.
Thus, for efficient characterisation, four specific slices were selected from the full 2D Rabi oscillation data and acquired following this procedure:
(1) Acquire a low-resolution 2D plot.
(2) Select an appropriate $\tau_\text{Rabi}$ and sweep the pump frequency (Fig.~\ref{fig:fitting}b).
(3) Once the frequency positions of Rabi 1--3 are identified, measure the Rabi signals by sweeping $\tau_\text{Rabi}$ (Fig.~\ref{fig:fitting}c,d).

During data fitting, the Hamiltonian parameters were determined iteratively in the following order:
(1) $P$ is primarily determined by the period of Rabi 1.
(2) $K$ is primarily determined by the frequency separation between the Rabi 1 and 2 positions.
(3) $\eta$ is primarily determined by the periods of Rabi 2 and 3 (Fig.~\ref{fig:fitting}c,d).
(4) $3\omega_\mathrm{K}$ is subsequently determined by matching the overall frequency positions of the Rabi 1--3 signals, as shown in Fig.~\ref{fig:fitting}b.
(5) These steps are iterated systematically via a grid search, sweeping $P$ in steps of $0.005$~MHz, $K$ in steps of $0.01$~MHz, and $\eta$ in steps of $0.005$.
A satisfactory fit is identified by visually matching essential dynamical features---specifically, the frequency separation and the periods of the Rabi signals---between the simulation and the measurement.

\subsection{Reconstruction of density matrices}

The Wigner function is defined as the expectation value of the photon number parity operator, $\hat{P}=\exp(\mathrm{i}\pi \hat{a}^\dagger\hat{a})$, at the phase-space coordinate $\alpha$ \cite{royer1977}:
\begin{equation} \label{eq:wigner}
W(\alpha) = \frac{2}{\pi} \mathrm{Tr}\mleft[\hat{D}^\dagger(\alpha) \rho \hat{D}(\alpha) \hat{P}\mright],
\end{equation}
where $\hat{D}(\alpha)=\exp(\alpha \hat{a}^\dagger - \alpha^* \hat{a})$ is the displacement operator and $\rho$ is the density matrix.
In this work, we reconstruct the unprocessed density matrix, $\rho_\textrm{CGAN}$, directly from the measured Wigner functions using quantum state tomography based on a conditional generative adversarial network (QST-CGAN).
For this implementation, we used the Python codes available at Ref.~\cite{qstcganCode}, which form the basis of the method detailed in Ref.~\cite{ahmed2021a}.

During the Wigner tomography, the physical displacement operation in the KPO is governed by
\begin{equation} \label{eq:dispOp}
\hat{D}_\textrm{K} = \exp\!\left\{\!
-\frac{\textrm{i}}{\hbar} \int_{0}^{\tau_\textrm{disp}}dt'
\left[ \beta(t')(\hat{a}^\dagger+\hat{a}) 
-\frac{K}{2}\hat{a}^\dagger\hat{a}^\dagger\hat{a}\hat{a} \right]
\! \right\},
\end{equation}
where $\tau_\textrm{disp}$ is the duration of the displacement pulse.
The presence of the self-Kerr term ($K$) introduces unwanted evolution that distorts the Wigner tomography.
Assuming the displacement pulse is sufficiently short, the dynamics induced by the displacement and Kerr terms can be treated as approximately separable.

Consequently, we can mitigate this distortion by applying a time-independent correction operator, $\hat{U}_\textrm{cor}$, to the reconstructed density matrix, such that $\rho = \hat{U}_\textrm{cor} \rho_\textrm{CGAN} \hat{U}_\textrm{cor}^\dagger$.We define this Kerr correction operator as
\begin{equation}\label{eq:cor}
\hat{U}_\textrm{cor}
= \exp(-\frac{\mathrm{i}}{\hbar} \frac{Kt_\textrm{cor}}{2} \hat{a}^\dagger\hat{a}^\dagger\hat{a}\hat{a}),
\end{equation}
where $t_\textrm{cor}$ is the effective Kerr correction time.
Note that the sign of the exponent is opposite to that in Eq.~\eqref{eq:dispOp} to invert the unwanted Kerr evolution.

In our previous work \cite{catGen}, we established that this Kerr distortion is reliably correctable (yielding a fidelity $\approx\! 1$) if the duration of a Gaussian displacement pulse (in nanoseconds) satisfies $\tau_\textrm{disp} < 1/[20\, (K/2\pi)]$, where $K/2\pi$ is in GHz.
For the experiment presented here, the Kerr coefficient at zero-pump amplitude is $K/2\pi = 1.70$ MHz.
Our measurement utilises a Gaussian displacement pulse with a length of $\tau_\textrm{disp}=14.0$ ns and a standard deviation of $\sigma=\tau_\textrm{disp}/6$, which is well within the correctable regime (the theoretical threshold is 29.4 ns).
Based on numerical simulations with these specific pulse parameters, we determined the correction time to be $t_\textrm{cor}=5.2$ ns, which is consistently applied to all density matrix reconstructions in this work.
Further details regarding the reliability of the Kerr correction protocol and the determination of $t_\textrm{cor}$ can be found in Ref.~\cite{catGen}.

We note that the validation of our QST-CGAN state reconstruction, including a cross-check against photon-number-splitting measurements, has been extensively demonstrated in our previous work \cite{catGen}.

In estimating the fidelity via QST-CGAN, the statistical uncertainties arising from measurement noise are evaluated to be less than $0.007$.
To obtain these conservative error bounds, we employed a Monte Carlo bootstrapping method.
We first extracted the standard deviation of the background noise from the edge pixels of the measured Wigner function.
We then generated artificial Gaussian noise with the identical deviation and injected it into the original measurement data.
By repeatedly reconstructing the quantum state and estimating the fidelity via QST-CGAN across these noisy pseudo-datasets, we obtained the final uncertainties.

\bigskip

\textbf{Data availability:} All data are available in the main text or in the supplementary information.


\bigskip

\textbf{Acknowledgments:}
The authors thank 
Akiyoshi Tomonaga of the National Institute of Advanced Industrial Science and Technology for providing niobium films,
the Semiconductor Science Research Support Team in RIKEN for technical support on fabrication,
and the MIT Lincoln Laboratory for providing a Josephson travelling-wave parametric amplifier.

\medskip

\textbf{Funding:}
This work was supported by the Japan Science and Technology Agency (Moonshot R\&D, JPMJMS2067) and the New Energy and Industrial Technology Development Organization (NEDO, JPNP16007).
SK acknowledges support from the National Research Foundation of Korea (NRF) grant (RS-2026-25472054, RS-2025-25463492) and the DGIST Start-up Fund Program (2026010391) funded by the Ministry of Science and ICT.

\medskip

\textbf{Author contributions:}
SK and JST conceived the project.
SK, DH, TN, HM, and JST designed the details of the experiment.
SK performed the measurements and data analysis with contributions from DH, TN, KT, and RK.
SW provided theoretical support.
DH and KT wrote the software for the measurements.
HM managed the hardware.
YZ and HM designed and prepared the sample package.
SK designed the chip.
TN, DS, and RK fabricated the chip.
SK wrote the original draft.
All authors contributed to the review and editing of the paper.
SK, FY, and JST supervised the project.
JST secured the majority of the funding.

\medskip

\textbf{Competing interests:} 
The authors declare that they have no competing interests.

\balancecolsandclearpage
\includepdf[pages=1]{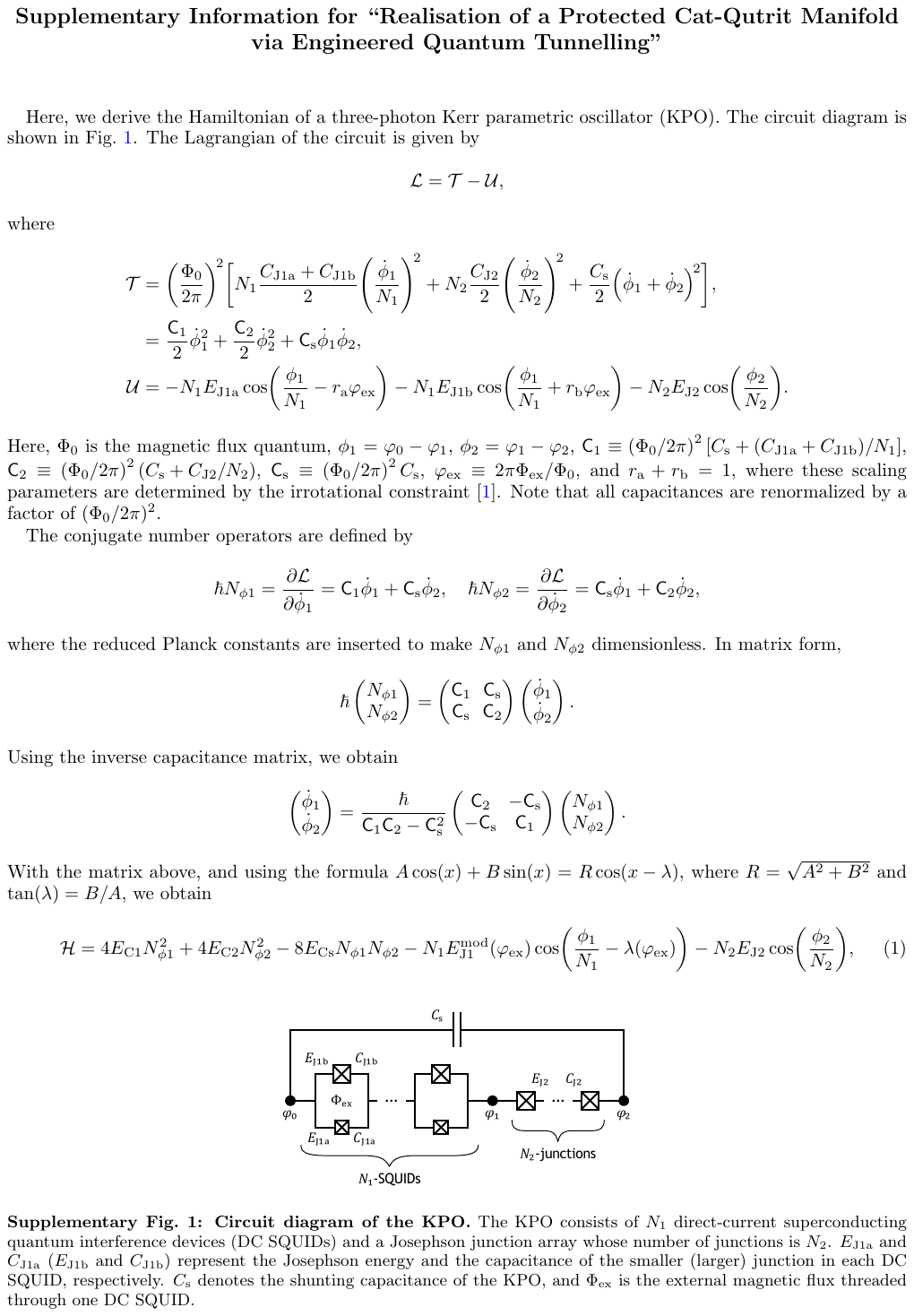}
$ $

\newpage
\includepdf[pages=2]{threePhotonCat_supp.pdf}
$ $
\newpage
\includepdf[pages=3]{threePhotonCat_supp.pdf}
$ $
\newpage
\includepdf[pages=4]{threePhotonCat_supp.pdf}
$ $
\newpage
\includepdf[pages=5]{threePhotonCat_supp.pdf}
$ $
\newpage
\includepdf[pages=6]{threePhotonCat_supp.pdf}
$ $

\end{document}